\documentclass[aps,twocolumn,showpacs]{revtex4}
\usepackage[dvips]{graphicx}
\usepackage{amsmath}
\newcommand{\bea}{\begin{eqnarray}}
\newcommand{\eea}{\end{eqnarray}}
\newcommand{\be}{\begin{equation}}
\newcommand{\ee}{\end{equation}}

\renewcommand{\a}{\mbox{$\alpha$}}

\newcommand{\BSCCO}{{Bi$_2$Sr$_2$CaCu$_2$O$_8$ }}

\def\k{{\bf k}}

\begin{document}

\title{Local electronic structure near oxygen dopants in
BSCCO-2212: \\a window on the high-$T_c$ pair mechanism? }

\author{Y.  He, T. S. Nunner$^{\dagger}$,  P. J. Hirschfeld$^*$, and H.-P. Cheng,
}

\affiliation{ Department of Physics, University of Florida, PO Box
118440, Gainesville FL 32611 USA\\$^\dagger$also Institut f\"ur
Theoretische Physik, Freie Universit\"at Berlin,
 Arnimallee 14, 14195 Berlin, Germany\\$^*$also
Laboratoire de Physique des Solides, Universit\'e Paris-Sud, 91405
Orsay, France }
\date{\today}

\begin{abstract}   The cuprate material \BSCCO (BSCCO-2212) is believed to be
doped by a combination of cation switching and excess oxygen. The
interstitial oxygen dopants are of particular interest because
scanning tunnelling microscopy (STM) experiments have shown that
they are positively correlated with the local value of the
superconducting gap, and calculations suggest that the fundamental
attraction between electrons is modulated locally. In this work,
we use density functional theory to try to ascertain which
locations in the crystal are energetically most favorable for the
O dopant atoms, and how the surrounding cage of atoms deforms.
 Our results provide support for the identification of STM
 resonances at -1eV with dopant interstitial O atoms, and show
 how the local electronic structure is modified nearby.

\end{abstract}

 \maketitle

%
STM measurements of impurities and other inhomogeneities in the
cuprates have opened a new window on  high temperature
superconductivity and raised a host of new questions about the way
these impurities interact with their
environment\cite{yazdani,davisnative,davisZn}. It has been
traditionally assumed that an impurity acts as a localized
screened Coulomb potential, and that its principal effect is the
 modification of quasiparticle wavefunctions nearby.  This effect
 is large and observable as a consequence of the $d$-wave symmetry
 of the superconducting state\cite{BZV05}.
Within such models, the  superconducting order parameter is
suppressed around the impurity, but this phenomenon is not usually
essential for qualitative predictions. Recently, a
phenomenological analysis\cite{NAMHprl05} of STM experiments
imaging interstitial oxygen atoms\cite{DavisScience05} suggested
that such impurities might have a much more striking effect, to
wit, the local modulation of the electronic pair interaction,
leading to large amplitude modulations of the superconducting
order parameter. If this is true, an understanding of the local
changes in electronic structure and couplings to collective
excitations of the material might tell us which aspects are of
crucial importance to  the pairing interaction itself.

Naturally occurring  impurities  which  dope the CuO$_2$ planes in
as-grown \BSCCO crystals\cite{Eisaki} include both the roughly 3\%
excess Bi substituting on the Sr sites, as well as oxygen
interstitials, whose concentration depends on the annealing
sequence and determines the  net doping of the sample. Both sets
of dopant atoms have recently been imaged by
STM\cite{Kinoda05,DavisScience05}. In the case of the O
interstitials, which appear as bright spots on the \BSCCO surface
at a bias of -960mV, a remarkable set of correlations was
established by McElroy et al.\cite{DavisScience05} between the
positions of the impurities and the magnitude of the local
superconducting gap (as defined by the position of the coherence
peak in the conductance signal), as well as the magnitude of the
LDOS at different energies. Local doping disorder was therefore
claimed to be responsible for the nanoscale inhomogeneities
observed earlier in the same
material\cite{cren,davisinhom1,davisinhom2,Kapitulnik1}, but the
{\it positive} dopant-gap correlation observed contradicted
earlier theoretical proposals along these
lines\cite{QHWang,balatskyLDC,Boston02}. Nunner et
al.\cite{NAMHprl05} then proposed that the oxygen dopant increased
the pairing strength $g$ between electrons locally, and a simple
model based on this ansatz was shown to explain all experimentally
reported correlations at optimal doping.  The results suggested
that the range of the modulation was of order a Cu-Cu distance.
While such atomic-scale enhancements of pairing are outside the
framework of standard BCS-type theoretical treatments of
impurities in superconductors, they have occasionally been
discussed in earlier treatments of the effects of impurities and
twin boundaries on conventional superconductors\cite{Earlytau1}.

If the theoretical proposal is correct, understanding where the O
interstitials sit and how they distort the cage of atoms around
them could be an important clue to the ingredients of the
electronic pair mechanism in these materials. Even if not, the
details of the local chemistry around interstitials can be
important to understand the doping process in the cuprates.  At
present, there is no consensus as to the exact position of the
defects imaged by STM, their chemical identity, nor  the nature of
the states they induce at -960 meV into which one tunnels in the
experiment.  We have therefore undertaken density functional
calculations of BSCCO-2212 including O atoms placed in the most
natural interstitial locations between the BiO and SrO, or between
the SrO and CuO$_2$ planes. We regard locations between other
layers as unlikely, since they are too far from the BiO layer,
where the BSCCO-2212 crystal typically cleaves, to be the source
of the strong STM signal observed. They are allowed to relax to
lower the overall energy of the system, and ultimately several
possible stable and metastable positions are found.  Only one of
these, close to but not identical to the position found by
statistical analysis of the STM data, represents a clear minimum
of the energy. We analyze the structural and electronic changes
thereby produced in some detail.

The most important results include a displacement of the nearby
apical (stoichiometric) oxygens from their naive positions on the
Bi-Cu axis and the creation of states centered primarily around
-1eV, not only associated with the interstitial O itself, but with
the apical oxygens and nearby Cu sites. It is the unhybridized
dopant O2$p_z$ states which give rise to the resonance seen by
STM. It is remarkable that an ab initio calculation in an
extremely complicated material is able to so quantitatively
reproduce the energetics of such  states, and we believe the
prospects are good, in light of this success, for wider
applicability of the technique to study other impurity problems in
the cuprates.

{\it Method.}
    In this work, DFT calculations have been used to determine all structural,
     energetic and electronic results. The Kohn-Sham equations are solved
     self-consistently in a plane-wave basis set, in conjunction with Vanderbilt
     ultrasoft pseudopotentials\cite{Vanderbilt90,Kresse}, which describe the electron-ion
     interaction,
     as implemented in the Vienna ab initio simulation program (VASP)\cite{Kresse96etseq}.
     Exchange
     and correlation are described by LDA. We use the exchange-correlation functional
     determined by Ceperly and Alder\cite{Ceperley80} and parameterized by Perdew and
     Zunger\cite{Perdew81}.  Details of the application of the
     method to impurity problems in \BSCCO have been given in
     \cite{WHC04}.  It is known that LDA fails  to give the
     correct band structure for cuprates in general, and in
     particular fails to describe the metallic/insulating
     character of the parent compounds.  On the other hand, the
     bands and Fermi surface given by LDA agree relatively well
     with ARPES results on  optimally doped BSCCO-2212, with minor
     exceptions (see below).
     Furthermore, as argued in \cite{WHC04}, we are primarily
     interested in high-energy, localized impurity states which should be
     little affected by the strong electronic correlations which renormalize the band states
      near the Fermi
     surface.

\begin{table}[t] \vskip .2cm
\begin{tabular}{|c|c|c|}\hline
  \# &Location (\AA) & Energy (eV)  \\
  \hline
  1 & 1.09,1.09,-0.93 & -6.56 \\
  2 & 2.11,0.00,-0.38 & -3.91
 \\
  3& 2.60,0.00, -0.38 & 4.38 \\
  4 & 2.60,0.00,-4.81& 14.97 \\
  \hline
\end{tabular}
\caption{Stable/metastable positions of O interstitial found.
Coordinates are relative to central Bi atom in Fig.
\ref{fig:pos1}. Energies are per O interstitial measured relative
to crystal without interstitial.} \label{table:positions}
\end{table}

   The BSCCO-2212
    surface is nearly always exposed at the cleaved BiO layer. For the calculation
    of the BSCCO surface in the presence of  impurities, we therefore use
    surface unit cells (super cells)
    with dimension of  $(2\sqrt{2}a\times 2\sqrt{2}a)$R45$^\circ$ that
    contain 8 primitive unit cells in the xy
    plane. There is one O interstitial added to the stoichiometric sample  of 120 atoms. The vacuum between
    neighboring slabs is about 15 \AA ~thick.

\begin{figure}[t]
\begin{center}
\includegraphics[width=1.\columnwidth]{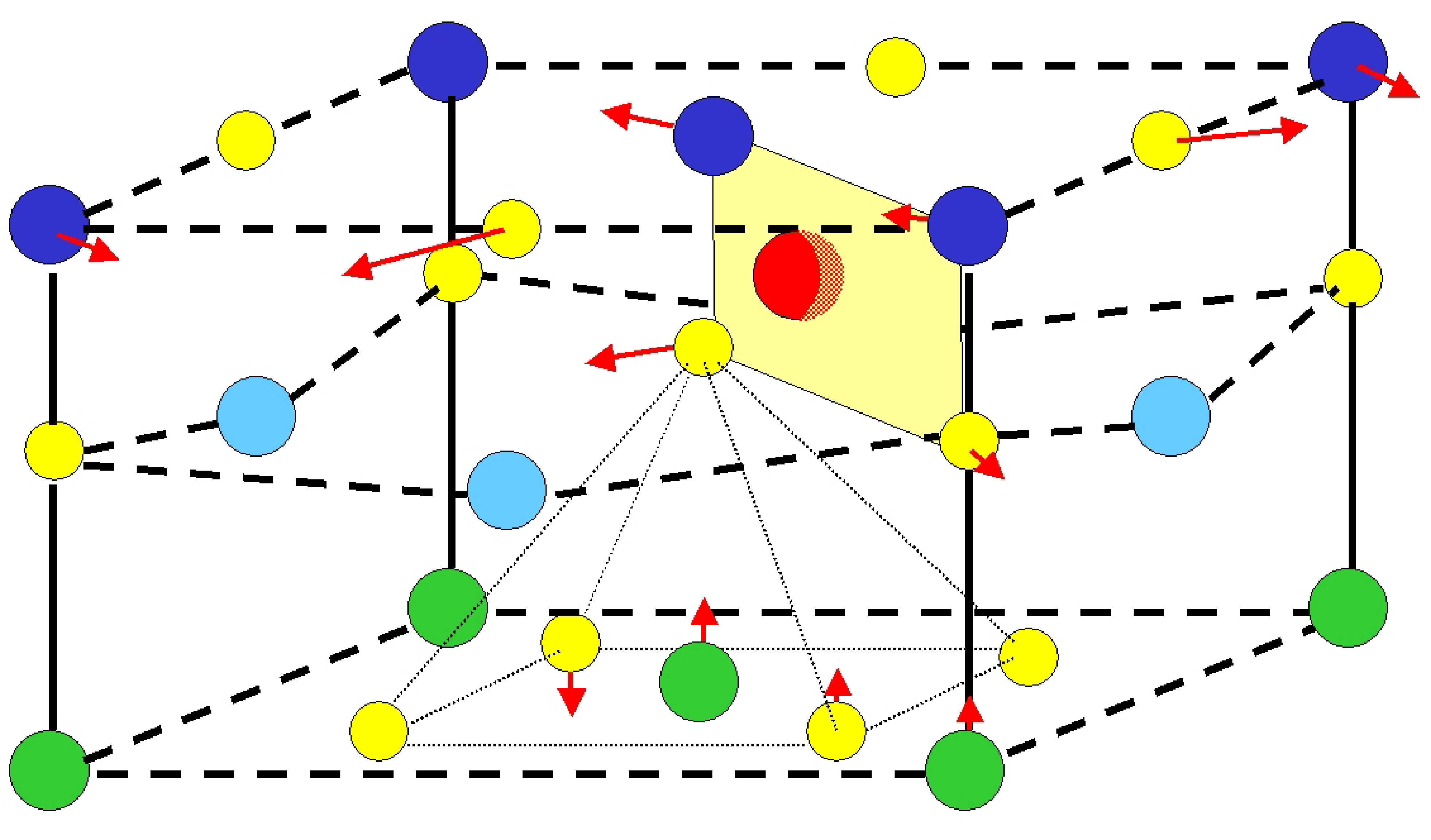}
\end{center}
\caption{(Color online) Atomic displacements due to the presence
of O interstitial at position 1 in table \ref{table:positions}.
Atomic species are color coded: Bi (blue), stoichiometric
O(yellow), Cu (green), Sr(light blue), Ca(aquamarine), O
interstitial (red).
 Arrows indicate rough measure of
displacement of atoms from their positions without interstitial.}
\label{fig:pos1}
\end{figure}

{\it  Structural relaxation.}  We  compute the structure and
electronic wavefunctions for the BSCCO-2212 surface first without
and then with an O interstitial.  In the latter case, the O is
positioned somewhere in the lattice and the entire system is
allowed to relax to minimize the total energy.   Several
metastable positions were found, summarized in Table
\ref{table:positions}.  It is seen that only one minimum thus
found is a plausible location (position 1 in the table) for such a
defect; others are higher in energy by at least 2.5 eV. Position 1
is located in the plane between 2 Bi atoms in the BiO layer and 2
stoichiometric oxygens in the SrO layer below, as shown in Fig.
\ref{fig:pos1}.  This position is displaced by about 0.5\AA ~and
has a different symmetry from that determined experimentally by
the STM group\cite{DavisScience05} by doing a statistical analysis
of their -960meV signal. This may be attributable to a combination
of dopant clustering effects and the distortions due to the
incommensurate  supermodulation in this material.

 The
displacements of atoms near the O impurity in position 1 are also
shown schematically in
    Figure \ref{fig:pos1}, and summarized in Table \ref{table:displacements}.
    The largest displacements are in the
    BiO and SrO planes.  In particular, we note that the so-called apical
    oxygen in the SrO plane is moved by nearly half an \AA~ away from the
    Bi-Cu axis, leading to an effective tilting of the
    semi-octahedron of stoichiometric oxygens away from this
    axis. Crudely speaking, this tilting is of the LTT
    (low-$T$ tetragonal)
    type, which has been associated with the tendency towards
    stripe formation in the underdoped
    cuprates\cite{KampfScalapino}.
 Finally, the Bi and O(Bi) atoms nearest the impurity in the
    BiO layer are also strongly pushed away by displacements of
    the same order.

\begin{table}[t]

\begin{tabular}{|c|c|c|c|}\hline
  Atom &${\bf r}_i$(\AA) &${\bf r}_f-{\bf r}_i$~~(\AA)&$|{\bf r}_f-{\bf r}_i|$~~(\AA)\\\hline
  Bi & 0,0,0  & -0.34,-0.34,-0.14 &  0.50\\
    Bi$^\prime$  & 2.59,2.59,0  & -0.17,-0.17,-0.08 & 0.25 \\
  O(Bi) & 2.59,0,0.05   & 0.68,-0.31,0.02 & 0.75 \\
  Sr & 2.59,0,-2.76&0.0,0.0,0.10  &    0.10  \\
  O(Sr) &0,0,-2.05& -0.41,-0.41,-0.12   &  0.59  \\
   O(Sr)$^\prime$ & 2.59,2.59,-2.05&0.16,0.16,-0.02    &  0.23  \\
  Cu &0,0,-4.60& -0.02,-0.02,0.14 &  0.14    \\\hline
\end{tabular}
\caption{Initial positions ${\bf r}_i$,  displacements ${\bf
r}_f-{\bf r}_i$, and displacement magnitude of various nearby
atoms in the presence of interstitial O at position 1 (see Table
1). Atoms are those of given species closest to O interstitial,
unless indicated by a $^\prime$ (second
closest).}\label{table:displacements}
\end{table}

{\it Changes in electronic structure.} Within the same framework,
we now calculate the electronic structure and changes thereto
caused by the O interstitials.  To compare directly to the STM
measurements, we should calculate the local density of states
(LDOS) a tip height distance ($\sim$ 3.5-4\AA) above the BiO
plane.  At present our numerical resolution is insufficient to
address this question directly, but we can ask in which energy
ranges new states are found, and roughly where in space they are
localized. The partial density of states (PDOS) projected onto
atomic species $\mu$ is defined as
\begin{eqnarray}
  \rho_\mu(\epsilon) &=& \sum_\k w_\k \sum_i
  \delta(\epsilon-\epsilon_{i,\k})|\langle
  \phi_{\mu,\k}|\Psi_{i,\k}\rangle|^2,\label{PDOS}
\end{eqnarray}
where $\phi_{\mu,\k}$ are linear combinations of atomic
wavefunctions consistent with the crystalline symmetry, and the
$\Psi_{i,\k}$ are Bloch wavefunctions for the $i$th band.  Here
$w_\k$ is the weight of each $\k$ point and $\epsilon_{i,\k}$ is
the dispersion of the Bloch state.

\begin{figure}[h]
\begin{center}
\includegraphics[width=.95\columnwidth,clip=true]{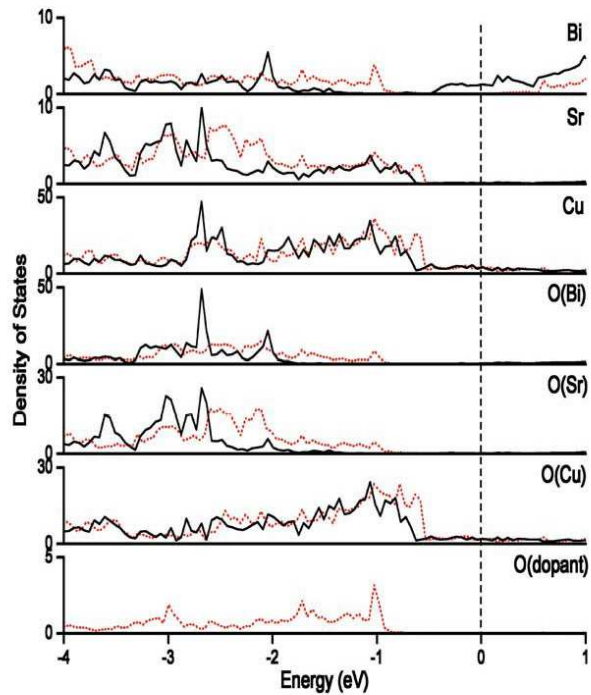}
\end{center}
\caption{ (Color online) Projected densities of states from Eq.
\ref{PDOS} for different atomic species before (black solid line)
and after (red dashed line) addition of O interstitial.}
\label{fig:PDOS}
\end{figure}

 In Figure \ref{fig:PDOS}, we plot the PDOS for relevant atomic
 orbitals in the absence of the O interstitial, together with the
 same quantities with the interstitial present. The spectra have been plotted relative
 to the absolute Fermi levels of 0.6527 eV and 0.4832 eV for the
 case without and with O interstitials, respectively.  The PDOS and DOS for the
 crystal without interstitials are consistent with earlier LDA calculations for
 this material\cite{KrakauerPickett}.   Upon addition of the O interstitial,
 one obvious change is the shift of the Cu and O(Cu) spectra towards the
 Fermi level, consistent with the hole doping of the CuO$_2$ plane
 by the interstitials.   Another
 interesting feature is the shift of PDOS Fermi level weight on the Bi
 sites to above the Fermi level upon doping with O. All previous LDA-type calculations of
 stoichiometric Bi-2212 have found BiO-derived bands crossing the Fermi
 level, leading to BiO Fermi surface ``pockets"\cite{Massidda}.
 These features have never been observed by
 ARPES\cite{DamascelliRMP}, suggesting that the doping by excess O
 may be necessary to obtain the correct Fermi
 surface\cite{Markiewicz}.

The primary effect of the interstitial O appears to be to
contribute states in a range around -1 to -3.5 eV, with a sharp
concentration of states around -1.0 eV itself. Since there are
originally no Bi, O(Bi), or O(Sr) states in this range, the extra
O apparently causes a significant change to the PDOS of these
species, inducing peaks at -1.0 eV in all three cases.  O
interstitial states in the rest of the energy regions do not
appear to induce qualitative changes in the PDOS of atomic species
near the surface, and may therefore be expected to not cause
significant qualitative changes to the LDOS measured by STM.  The
only expected strong signal of the presence of the O defects is
therefore a concentration of (assumed) tunnelling states at -1.0
eV, nearly precisely where such a signal is found by
STM\cite{DavisScience05}.  This is a spectacular confirmation of
the ability of DFT to predict electronic structure  near
impurities with high accuracy in these systems. We note that the
rough magnitude of the binding energy of these states can be
understood by recognizing the similarity to the {\it
stoichiometric} O2$p$ derived states (found at $\sim$ -1.5eV below
the top of the Zhang-Rice band) in photoemission experiments on
Sr$_2$CuO$_2$Cl$_2$, which do not hybridize with Cu 3d
states\cite{Sawatzky}.  The peaks around -1eV seen in the O(Cu)
and Cu spectra in Figure 2 which do not change significantly with
the addition of O dopants are  of this type.

We note finally the existence of a broad cluster of O(dopant)
states centered around -1.7 eV which also appear to induce similar
features in O(Bi) and Bi, and could lead to significant LDOS
change at the BSCCO surface.  This is an independent prediction
which can be checked by STM.

\vskip .2cm ~

\begin{figure}[h]
\begin{center}
\includegraphics[width=\columnwidth]{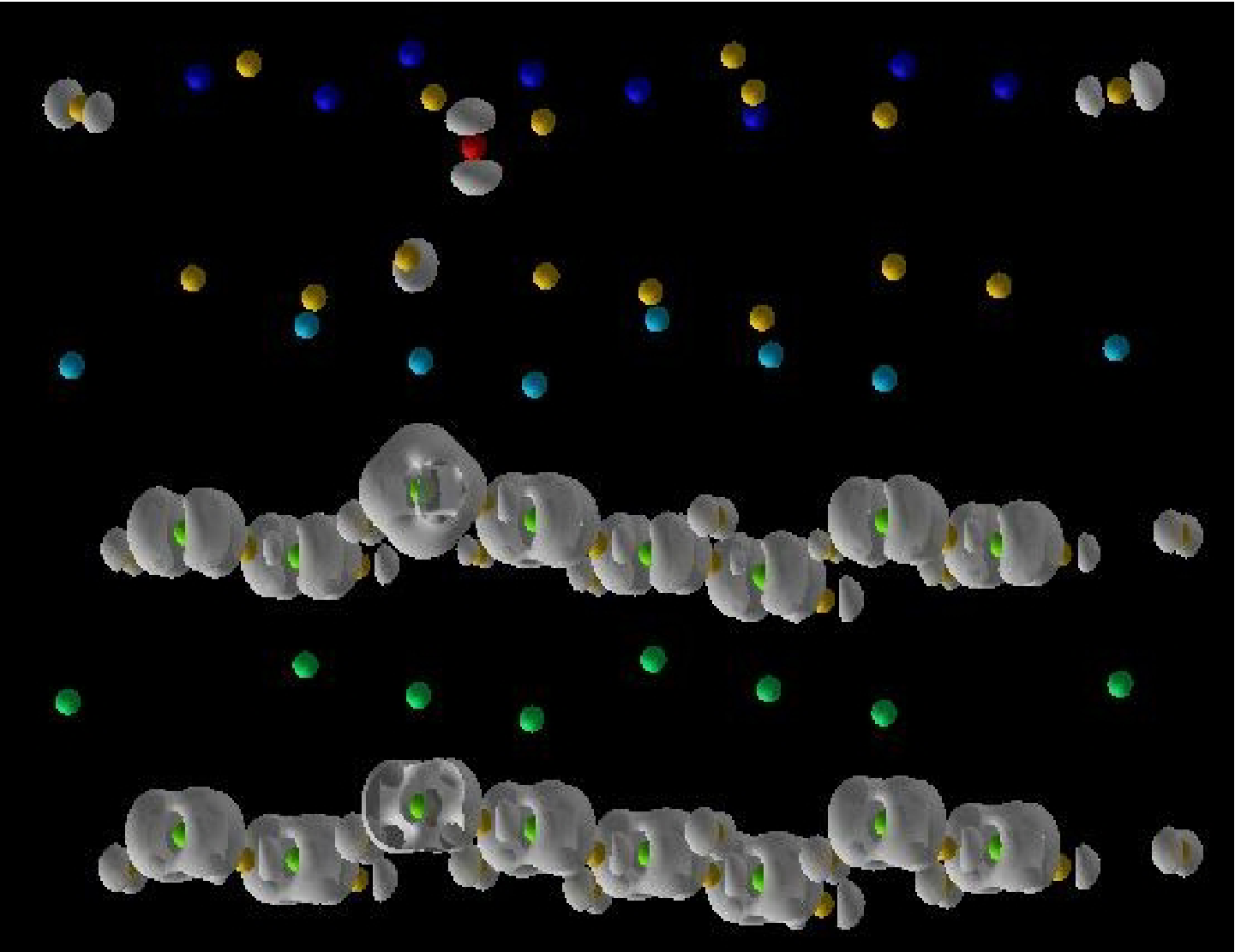}
\end{center}
\caption{(Color online) Spatial distribution of LDOS integrating
over states near -1eV.  The surface of the sample is at the top of
the figure.
 Atomic species are color coded: Bi (blue), O(yellow), Cu (green), Sr(light blue),
 Ca(aquamarine), O interstitial (red).
} \label{fig:LDOS}
\end{figure}

{\it Local density of states near -1eV.} The PDOS is calculated
for all atoms of a particular species and symmetry, but spatially
resolved LDOS information is necessary to specify where these
changes are taking place relative to the interstitial site. To
this end, we focus on the states near -1 eV, and plot a contour of
the LDOS

\begin{eqnarray}
  \rho(\epsilon,{\bf r}) &=& \sum_\k w_\k\sum_i
  \delta(\epsilon-\epsilon_{i,\k}) |\Psi_{i,\k}({\bf r})|^2
\end{eqnarray}
 integrated over a range of
energies from -1.1eV to -0.9eV   in Figure \ref{fig:LDOS}.
Inspection of various contours of this type leads to the
conclusion that the states at -1eV are dominated by  unhybridized
CuO$_2$ plane states\cite{Sawatzky}, {\it and} induced states
localized near the impurity.
In particular, the slightly misoriented O2$p_z$ state on the O
interstitial is clearly visible. There is no other state in the
vicinity close to the surface positioned and oriented to
accomodate an electron transfer from an STM tip near the surface.
We note the absence of any states localized on the Bi atoms, in
apparent contradiction to the results shown in Fig.
\ref{fig:PDOS}.  This is  because the Bi radius is 2 times larger,
such that the PDOS weight appears at smaller LDOS values than
depicted here.
\\ \indent
The next fascinating change due to the O interstitial is the
creation of new states on the closest Cu site which appear to be
of $3d_{3z^2-1}$ symmetry. This process is clearly mediated by the
apical O(Sr), which also shows induced 2$p_{x,y}$ symmetry states
in this range. There is no obvious hybridization between these
states; rather, these appear to be changes due to the
displacements of the atoms in question, which transfer spectral
weight from higher binding energy into this range, as seen in Fig.
\ref{fig:PDOS}. \\ \indent
 {\it Conclusions.} Using density
functional theory methods, we have shown that the excess O dopant
atoms in the as grown BSCCO-2212 material are located with high
probability between the BiO and SrO layers in the plane containing
Bi and O(Sr) atoms. These interstitials are responsible for the
creation of a set of states narrowly localized near -1eV, which we
have investigated more closely.   One unhybridized 2$p$ state
appears on the dopant itself, and is oriented in the $z$
direction, strongly suggesting that this is precisely the state
imaged in recent STM experiments \cite{DavisScience05} at -960
meV.  This result is a strong confirmation of the ability of DFT
to calculate high-energy impurity features even in strongly
correlated systems, as emphasized in Ref. \cite{WHC04}. In
addition, new states in this range are induced on the strongly
perturbed nearby apical oxygen site, and even on the closest Cu
atom.  This suggests that the pairing of two electrons in the
CuO$_2$ plane in this high-temperature superconductor may indeed
be influenced by the presence of the nearby O interstitial, as
suggested in Ref. \cite{NAMHprl05}, and that information on the
modulation of the pair interaction may be gleaned by
``downfolding" the DFT information to determine how the local
change in electronic structure correlates with the superconducting
gap.
\\ \indent
 {\it Acknowledgements.} This work was supported by ONR
 N00014-04-0060 (PJH), DOE grants DE-FG02-05ER46236 (PJH) and DE-FG02-97ER45660 (H-PC), and
 NSF/DMR/ITR-medium program under contract number DMR-032553 (H-PC), and the A. v. Humboldt Foundation
 (TSN). The authors
 acknowledge valuable discussions
 with B.M. Andersen, W.A. Atkinson,
 J.C. Davis, T. Devereaux, A. Kampf, A. Melikyan, G. A. Sawatzky and L.-L. Wang.

\end{document}